\documentclass[12pt]{article}
\usepackage[latin9]{inputenc}
\usepackage[active]{srcltx}
\usepackage{amsmath}
\usepackage{amssymb}
\usepackage[pdfusetitle,
 bookmarks=true,bookmarksnumbered=false,bookmarksopen=false,
 breaklinks=false,pdfborder={0 0 0},pdfborderstyle={},backref=false,colorlinks=false]
 {hyperref}

\makeatletter

\DeclareTextSymbolDefault{\textquotedbl}{T1}

\usepackage{amsfonts}
\usepackage{ytableau} 
\usepackage[square,numbers,sort&compress]{natbib}

\makeatother

\begin{document}
\begin{flushright}
FIAN/TD/12-2024
\par\end{flushright}

\vspace{0.5cm}
 
\begin{center}
{\large\textbf{Unfolded Formulation of $4d$ Yang\textendash Mills
Theory}}{\large\par}
\par\end{center}

\begin{center}
\par\end{center}

\begin{center}
\vspace{0.2cm}
 \textbf{Nikita Misuna}\\
 \vspace{0.5cm}
 \textit{Tamm Department of Theoretical Physics, Lebedev Physical
Institute,}\\
 \textit{Leninsky prospekt 53, 119991, Moscow, Russia}\\
 
\par\end{center}

\begin{center}
\vspace{0.6cm}
 misuna@lpi.ru \\
 
\par\end{center}

\vspace{0.4cm}

\begin{abstract}
In this note, we present a novel formulation of $4d$ pure Yang-Mills
theory within the unfolded framework of Vasiliev higher-spin gravity.
This formulation is first-order and exhibits manifest diffeomorphism
and gauge invariance. Our approach builds upon a recently proposed
unfolding method, previously applied to scalar electrodynamics. Additionally,
we discuss the features of various unfolding maps defined by the unfolded
equations.

\newpage{}
\end{abstract}

\subsubsection*{Introduction}

Symmetry considerations serve as one of the most fundamental guiding
principles in constructing theories of fundamental interactions. Higher-spin
gravity theories provide prime examples, describing systems of interacting
massless fields of all spins. To a large extent, these theories are
determined by an infinite-dimensional higher-spin gauge symmetry \cite{HSalgebra}.
For a partial overview of the relevant literature, see \cite{snow}.

Maintaining control over higher-spin symmetry is of fundamental importance.
The first example of higher-spin gravity was formulated through the
Vasiliev equations \cite{vas1,vas2}, a set of classical equations
of motion written in the so-called unfolded form. Unfolded equations
are first-order differential equations for unfolded fields \textendash{}
which are exterior forms \textendash{} that possess manifest gauge
symmetry. Thus, the unfolded dynamics approach \cite{unf1,vas1,vas2,unf2,unf3,ActionsCharges}
provides a classical first-order formalism that ensures both manifest
diffeomorphism and gauge invariance. It also enables effective control
over the theory's degrees of freedom: unfolded fields (which form
an infinite spectrum in dynamical field theories) parameterize all
d.o.f. of the system. In particular, these properties make the unfolded
dynamics approach a promising tool for investigating $AdS/CFT$ correspondence
and other dualities \cite{hologr1,hologr2,hologr3}. A quantization
scheme for classical unfolded theories was put forward in \cite{unfQFT}
(see also \cite{quant1,quant2,quant3,Grigoriev} for discussions of
quantization of non-Lagrangian field theories).

Beyond higher-spin gravity models \cite{vas1,vas2,ss1,ss2,did1,did2},
relatively few examples of unfolded formulations exist for nonlinear
theories \cite{unfQFT,qed} (see also \cite{unf_conf} for a discussion
of nonlinear unfolding in conformal geometry), thereby constraining
comprehensive exploration of the approach's full potential. The primary
obstacle was the lack of a practical unfolding algorithm \textendash{}
especially critical for nonlinear theories. Therefore, linear models
were mainly studied \cite{unf4,unf5,unf6,unf7,unf10}. In \cite{qed},
a novel unfolding method was put forward, that enabled the construction
of an unfolded formulation of $4d$ scalar electrodynamics. In essence,
it involves two steps: first postulating the form of an unfolded master-field,
then deriving the corresponding unfolded equations as identities for
this master-field.

In this note, we adapt and refine the method of \cite{qed} to construct
an unfolded formulation of $4d$ on-shell pure Yang\textendash Mills
theory.

The note is organized as follows. First, we review basic concepts
of the unfolded dynamics approach. Next, we construct the unfolded
Yang\textendash Mills equations by fixing the form of the unfolded
master-fields and using certain operator relations that we derive.
We then discuss the properties of the resulting unfolded system and
comment on the unfolding maps it defines. In conclusion, we outline
promising directions for future research.

\subsubsection*{Unfolded dynamics approach}

Within the unfolded dynamics approach \cite{unf1,vas1,vas2,unf2,unf3,ActionsCharges},
field theories are formulated via first-order differential equations
\begin{equation}
\mathrm{d}W^{A}(x)+G^{A}(W)=0\label{unf_eq}
\end{equation}
on unfolded fields $W^{A}(x)$, which are exterior forms on a space-time
manifold $M^{d}$. Here, $A$ collectively denotes all indices of
an unfolded field, $\mathrm{d}$ is the exterior derivative on $M^{d}$,
and $G^{A}(W)$ is constructed from exterior products of unfolded
fields (the wedge symbol is omitted throughout the paper). For each
$W^{A}$ there is one and only one unfolded equation \eqref{unf_eq}. 

The system \eqref{unf_eq} must obey the consistency condition
\begin{equation}
G^{B}\dfrac{\delta G^{A}}{\delta W^{B}}\equiv0,\label{unf_consist}
\end{equation}
which follows from $\mathrm{d}^{2}\equiv0$ and constrains possible
forms of $G^{A}$. Considering this, the unfolded equations \eqref{unf_eq}
possess manifest gauge symmetry\footnote{Rigorously, this holds provided the consistency is dimension-independent
\textendash{} a property satisfied by all known examples \cite{ActionsCharges}.}
\begin{equation}
\delta W^{A}=\mathrm{d}\varepsilon^{A}(x)-\varepsilon^{B}\dfrac{\delta G^{A}}{\delta W^{B}},\label{unf_gauge_transf}
\end{equation}
with each $(n>0)$-form field $W^{A}$ giving rise to a gauge symmetry
with $(n-1)$-form parameter $\varepsilon^{A}(x)$, while 0-form fields
transform only under gauge symmetries of 1-form fields through the
second term in \eqref{unf_gauge_transf}.

In dynamical field theories, the spectrum of unfolded fields is infinite,
as they parameterize all physical d.o.f. Typically, the space of unfolded
fields admits a grading bounded from below, so that equations \eqref{unf_eq}
express (perhaps, in a very complicated nonlinear way) higher-grade
fields through derivatives of lower-grade ones. In addition, \eqref{unf_eq}
may implicitly impose dynamical constraints on the lowest-grade fields,
rendering the system on-shell. Equations \eqref{unf_eq} define what
is known as a free differential algebra \cite{Sullivan} (see also
\cite{Barnich,Grigoriev} for discussions of the relation between
the unfolded framework and various mathematical structures).

Ultimately, an unfolded system \eqref{unf_eq} describes a theory
of lowest-grade fields (also referred to as primaries), possibly subject
to some differential constraints (e.o.m. or whatever), while higher-grade
fields constitute (infinite) towers of their covariant differential
descendants. Crucially, the unfolded formulation maintains manifest
diffeomorphism and gauge invariance and provides complete control
over d.o.f. of the theory. These features establish unfolded dynamics
approach as a powerful framework for investigating higher-spin gravity.
However, these advantages extend beyond higher spins, offering potential
applications to conventional field theories. In this note, our goal
is to reformulate $4d$ pure Yang\textendash Mills theory in the unfolded
form \eqref{unf_eq}, applying the unfolding method introduced in
\cite{qed}. 

\subsubsection*{Yang\textendash Mills fields and auxiliary spinors}

In $sl(2,\mathbb{C})$-spinor notation, the $4d$ Yang\textendash Mills
equations together with Bianchi identities are
\begin{equation}
\mathrm{D}_{\beta\dot{\alpha}}F^{\beta}{}_{\alpha}=0,\hspace*{1em}\mathrm{D}_{\alpha\dot{\beta}}\bar{F}^{\dot{\beta}}{}_{\dot{\alpha}}=0,\label{YM_eq_F}
\end{equation}
where the (anti-)self-dual components of the field strength tensor
are
\begin{equation}
F_{\alpha\alpha}:=\frac{\partial}{\partial x^{\alpha\dot{\beta}}}A_{\alpha}{}^{\dot{\beta}}-i[A_{\alpha\dot{\beta}},A_{\alpha}{}^{\dot{\beta}}],\hspace*{1em}\bar{F}_{\dot{\alpha}\dot{\alpha}}:=\frac{\partial}{\partial x^{\beta\dot{\alpha}}}A^{\beta}{}_{\dot{\alpha}}-i[A_{\beta\dot{\alpha}},A^{\beta}{}_{\dot{\alpha}}],\label{YM_tensor}
\end{equation}
and the covariant derivative is
\begin{equation}
\mathrm{D}_{\alpha\dot{\alpha}}:=\frac{\partial}{\partial x^{\alpha\dot{\alpha}}}-i[A_{\alpha\dot{\alpha}},\bullet],\label{covar_derivative_def}
\end{equation}
satisfying
\begin{equation}
[\mathrm{D}_{\alpha\dot{\alpha}},\mathrm{D}_{\beta\dot{\beta}}]=-i\epsilon_{\alpha\beta}\bar{F}_{\dot{\alpha}\dot{\beta}}-i\epsilon_{\dot{\alpha}\dot{\beta}}F_{\alpha\beta}.\label{covar_derivative_comm}
\end{equation}
The antisymmetric spinor metric
\begin{equation}
\epsilon_{\alpha\beta}=\epsilon_{\dot{\alpha}\dot{\beta}}=\left(\begin{array}{cc}
0 & 1\\
-1 & 0
\end{array}\right),\quad\epsilon^{\alpha\beta}=\epsilon^{\dot{\alpha}\dot{\beta}}=\left(\begin{array}{cc}
0 & 1\\
-1 & 0
\end{array}\right)
\end{equation}
raises/lowers spinor indices via
\begin{equation}
v_{\alpha}=\epsilon_{\beta\alpha}v^{\beta},\quad v^{\alpha}=\epsilon^{\alpha\beta}v_{\beta},\quad\bar{v}_{\dot{\alpha}}=\epsilon_{\dot{\beta}\dot{\alpha}}\bar{v}^{\dot{\beta}},\quad\bar{v}^{\dot{\alpha}}=\epsilon^{\dot{\alpha}\dot{\beta}}\bar{v}_{\dot{\beta}}.
\end{equation}
For multispinors, indices denoted with the same letter are either
contracted or symmetrized, depending on their relative positions,
(similarly for dotted indices)
\begin{equation}
T_{\alpha\alpha}:=T_{(\alpha_{1}\alpha_{2})},\quad T_{\alpha}{}^{\alpha}:=\epsilon^{\alpha\beta}T_{\alpha\beta}.
\end{equation}

In order to unfold Yang\textendash Mills theory, one has to introduce,
on top of primaries $F_{\alpha\alpha}$ and $\bar{F}_{\dot{\alpha}\dot{\alpha}}$,
the infinite towers of all their differential on-shell descendants.
This can be conveniently performed by means of auxiliary commuting
spinors\footnote{In $4d$ Vasiliev equations \cite{vas1,vas2}, these spinors become
generators of an associative higher-spin gauge algebra through a specific
non-commutative star product defined on them.} $Y=(y^{\alpha},\bar{y}^{\dot{\alpha}})$, then the whole towers get
packed into unfolded Yang\textendash Mills master-fields, which we
postulate to be of the form
\begin{equation}
F(Y|x)=e^{\mathrm{D}y\bar{y}}F_{\alpha\alpha}(x)y^{\alpha}y^{\alpha}e^{-\mathrm{D}y\bar{y}},\quad\bar{F}(Y|x)=e^{\mathrm{D}y\bar{y}}\bar{F}_{\dot{\alpha}\dot{\alpha}}(x)\bar{y}^{\dot{\alpha}}\bar{y}^{\dot{\alpha}}e^{-\mathrm{D}y\bar{y}},\label{YM_field_postulate}
\end{equation}
where right-acting derivatives are contracted with spinors as
\begin{equation}
\mathrm{D}y\bar{y}:=y^{\alpha}\bar{y}^{\dot{\alpha}}\mathrm{D}_{\alpha\dot{\alpha}}.
\end{equation}
Unfolded master-fields \eqref{YM_field_postulate} contain primary
Yang-Mills tensors as (anti-)holomorphic in $Y$ components
\begin{equation}
F_{\alpha\alpha}(x)y^{\alpha}y^{\alpha}=F(Y|x)|_{\bar{y}=0},\quad\bar{F}_{\dot{\alpha}\dot{\alpha}}(x)\bar{y}^{\dot{\alpha}}\bar{y}^{\dot{\alpha}}=\bar{F}(Y|x)|_{y=0},
\end{equation}
together with an infinite sequence of their fully symmetrized traceless
covariant derivatives of all orders. This constitutes the set of all
independent covariant descendants of the primary Yang\textendash Mills
tensor, since antisymmetrizations and contractions are determined
by \eqref{YM_eq_F} and \eqref{covar_derivative_comm}. By construction,
unfolded master-fields \eqref{YM_field_postulate} inherit the adjoint
representation of the gauge algebra of primary tensors.

The aforementioned grading on the space of unfolded fields can be
introduced in terms of the spinor Euler operators
\begin{equation}
N:=y^{\alpha}\partial_{\alpha},\quad\bar{N}:=\bar{y}^{\dot{\alpha}}\bar{\partial}_{\dot{\alpha}},\label{Euler_operators}
\end{equation}
where $\partial_{\alpha}$ and $\bar{\partial}_{\dot{\alpha}}$ are
$y^{\alpha}$- and $\bar{y}^{\dot{\alpha}}$-derivatives. From their
definitions \eqref{YM_field_postulate}, master-fields obey
\begin{equation}
(N-\bar{N})F=2F,\quad(N-\bar{N})\bar{F}=-2\bar{F},\label{helicity}
\end{equation}
which corresponds to helicities $\pm1$ (in higher-spin gravity, strength
tensors with $|N-\bar{N}|F=2sF$ describe spin-$s$ massless fields)
and, further,
\begin{equation}
\mathrm{D}y\bar{y}F=\bar{N}F,\quad\mathrm{D}y\bar{y}\bar{F}=N\bar{F}.\label{Dyy_F}
\end{equation}
Then a grading operator on the space of unfolded fields can be defined
as
\begin{equation}
\mathcal{G}:=\frac{1}{2}(N+\bar{N}-2).\label{grading_operator}
\end{equation}
It decomposes the space of master-fields $F$ into a direct sum of
its eigenspaces (the same applies to $\bar{F}$)
\begin{equation}
F(Y|x)=\sum_{n=0}^{\infty}F^{(n)}(Y|x),\quad\mathcal{G}F^{(n)}=nF^{(n)},
\end{equation}
so that the primary has $0$-grade, while $n$-grade unfolded fields
represent its covariant derivatives of $n$-th order, as follows from
\eqref{YM_field_postulate}.

To formulate the unfolded equations \eqref{unf_eq}, one needs to
express the derivatives of the unfolded fields in algebraic terms.
In our particular case, the task is to express $\mathrm{D}_{\alpha\dot{\beta}}F$
for $F$ \eqref{YM_field_postulate} in terms of $Y$ and $\partial/\partial Y$
acting on $F$ and $\bar{F}$ (and the same for $\mathrm{D}_{\alpha\dot{\beta}}\bar{F}$).
Before we start processing \eqref{YM_field_postulate}, let us work
out some general operator relations that simplify the analysis implied
by the method of \cite{qed}. Combining a commutator formula
\begin{equation}
[\hat{\mathrm{A}},e^{\hat{\mathrm{D}}}]=\int_{0}^{1}dte^{t\hat{\mathrm{D}}}[\hat{\mathrm{A}},\hat{\mathrm{D}}]e^{-t\hat{\mathrm{D}}}e^{\hat{\mathrm{D}}}\label{comm_1}
\end{equation}
with an Euler-operator representation of a homotopy integral
\begin{equation}
\int_{0}^{1}dtt^{k}F(tz)=\frac{1}{z\frac{\partial}{\partial z}+1+k}F(z),\label{homotopy}
\end{equation}
one obtains
\begin{equation}
[\hat{\mathrm{A}},e^{\hat{\mathrm{D}}}]=(\frac{1}{N_{\hat{\mathrm{D}}}}e^{\hat{\mathrm{D}}}[\hat{\mathrm{A}},\hat{\mathrm{D}}]e^{-\hat{\mathrm{D}}})e^{\hat{\mathrm{D}}}=e^{\hat{\mathrm{D}}}(\frac{1}{N_{\hat{\mathrm{D}}}}e^{-\hat{\mathrm{D}}}[\hat{\mathrm{A}},\hat{\mathrm{D}}]e^{\hat{\mathrm{D}}}),\label{comm_2}
\end{equation}
where the inverse Euler operator is understood (when $F(0)=0$) as
\begin{equation}
\frac{1}{N_{\hat{\mathrm{D}}}}F(\hat{\mathrm{D}}):=\int_{0}^{1}dt\frac{1}{t}F(t\hat{\mathrm{D}}).\label{Euler_homotopy}
\end{equation}
Applying this formula to an operator of the form
\begin{equation}
\hat{\mathrm{B}}=e^{\hat{\mathrm{D}}}\hat{\mathrm{C}}e^{-\hat{\mathrm{D}}}\label{B_def}
\end{equation}
yields
\begin{equation}
[\hat{\mathrm{A}},\hat{\mathrm{B}}]=[(\frac{1}{N_{\hat{\mathrm{D}}}}e^{\hat{\mathrm{D}}}[\hat{\mathrm{A}},\hat{\mathrm{D}}]e^{-\hat{\mathrm{D}}}),\hat{\mathrm{B}}]+e^{\hat{\mathrm{D}}}[\hat{\mathrm{A}},\hat{\mathrm{C}}]e^{-\hat{\mathrm{D}}}.\label{commutator_1}
\end{equation}
Next, considering the case $\hat{\mathrm{C}}=\hat{\mathrm{D}}$, from
\eqref{commutator_1} one gets
\begin{equation}
e^{\hat{\mathrm{D}}}[\hat{\mathrm{A}},\hat{\mathrm{D}}]e^{-\hat{\mathrm{D}}}=[\hat{\mathrm{A}},\hat{\mathrm{D}}]-\frac{1}{N_{\hat{\mathrm{D}}}-1}(e^{\hat{\mathrm{D}}}[[\hat{\mathrm{A}},\hat{\mathrm{D}}],\hat{\mathrm{D}}]e^{-\hat{\mathrm{D}}}),\label{comm_3}
\end{equation}
so that \eqref{commutator_1} can be equivalently rewritten as
\begin{equation}
[\hat{\mathrm{A}},\hat{\mathrm{B}}]=[(\frac{1}{N_{\hat{\mathrm{D}}}(N_{\hat{\mathrm{D}}}-1)}e^{\hat{\mathrm{D}}}[\hat{\mathrm{D}},[\hat{\mathrm{A}},\hat{\mathrm{D}}]]e^{-\hat{\mathrm{D}}}),\hat{\mathrm{B}}]+[[\hat{\mathrm{A}},\hat{\mathrm{D}}],\hat{\mathrm{B}}]+e^{\hat{\mathrm{D}}}[\hat{\mathrm{A}},\hat{\mathrm{C}}]e^{-\hat{\mathrm{D}}}.\label{commutator_2}
\end{equation}

For convenience, we denote an \textquotedbl unfolding map\textquotedbl{}
for an arbitrary field $\mathrm{C}_{\alpha(n),\dot{\beta}(m)}(Y|x)$
taking values in the adjoint representation as
\begin{equation}
\ll\mathrm{C}_{\alpha(n),\dot{\beta}(m)}(Y|x)\gg:=e^{\mathrm{D}y\bar{y}}\mathrm{C}_{\alpha(n),\dot{\beta}(m)}(Y|x)e^{-\mathrm{D}y\bar{y}},\label{unf_brackets}
\end{equation}
so that, in particular,
\begin{equation}
F(Y|x)=\ll F_{\alpha\alpha}(x)y^{\alpha}y^{\alpha}\gg,\quad\bar{F}(Y|x)=\ll\bar{F}_{\dot{\alpha}\dot{\alpha}}(x)\bar{y}^{\dot{\alpha}}\bar{y}^{\dot{\alpha}}\gg.
\end{equation}
Then from \eqref{commutator_2} one has (here $C\equiv\mathrm{C}_{\alpha(n),\dot{\beta}(m)}(Y|x)$
)
\begin{equation}
\ll\partial_{\mu}\mathrm{C}\gg=(\partial_{\mu}-\mathrm{D}_{\mu\dot{\mu}}\bar{y}^{\dot{\mu}})\ll\mathrm{C}\gg+iy_{\mu}[\frac{1}{(N+1)(N+2)}\bar{F},\ll\mathrm{C}\gg],\label{d_C}
\end{equation}
\begin{equation}
\ll\bar{\partial}_{\dot{\mu}}\mathrm{C}\gg=(\bar{\partial}_{\dot{\mu}}-\mathrm{D}_{\mu\dot{\mu}}y^{\mu})\ll\mathrm{C}\gg+i\bar{y}_{\dot{\mu}}[\frac{1}{(\bar{N}+1)(\bar{N}+2)}F,\ll\mathrm{C}\gg].\label{dbar_C}
\end{equation}
From here on, all derivatives inside the angle brackets always act
only on an expression within brackets and never differentiate unfolding
exponents of \eqref{unf_brackets}. The square brackets stand for
the commutator in the gauge Lie algebra and the Euler operators of
\eqref{commutator_2} are expressed in terms of the spinor Euler operators
\eqref{Euler_operators}. Master-fields $F$ and $\bar{F}$ arise
in \eqref{d_C}-\eqref{dbar_C} through \eqref{covar_derivative_comm}
without assuming Yang\textendash Mills equations \eqref{YM_eq_F}.

Now, let us get down to solving the problem. Direct application of
\eqref{commutator_1} to $\mathrm{D}_{\mu\dot{\mu}}F(Y|x)$ gives
\begin{equation}
\mathrm{D}_{\mu\dot{\mu}}F=[\frac{1}{N}\ll[\mathrm{D}_{\mu\dot{\mu}},\mathrm{D}y\bar{y}]\gg,F]+\ll\mathrm{D}_{\mu\dot{\mu}}F_{\alpha\alpha}y^{\alpha}y^{\alpha}\gg.\label{D_F_initial}
\end{equation}
The task is to eliminate all covariant derivatives and angle brackets
on the r.h.s. by re-expressing them in terms of unfolded master-fields
$F$ and $\bar{F}$ acted on by $Y$'s and $\frac{\partial}{\partial Y}$'s.
To accomplish this, we have the following tools at our disposal: Yang\textendash Mills
equations \eqref{YM_eq_F} together with \eqref{covar_derivative_comm},
the relations \eqref{d_C} and \eqref{dbar_C}, the Schouten identities
for spinors and the Jacobi identity of the gauge Lie algebra.

Applying \eqref{covar_derivative_comm} to the first term on the r.h.s.
of \eqref{D_F_initial} and \eqref{YM_eq_F} (plus Schouten identities)
to the second one, one has
\begin{equation}
\mathrm{D}_{\mu\dot{\mu}}F=\frac{i}{2}\bar{y}_{\dot{\mu}}[\frac{1}{N}\ll\partial_{\mu}F_{\alpha\alpha}y^{\alpha}y^{\alpha}\gg,F]+\frac{i}{2}y_{\mu}[\frac{1}{\bar{N}}\ll\bar{\partial}_{\dot{\mu}}\bar{F}_{\dot{\alpha}\dot{\alpha}}\bar{y}^{\dot{\alpha}}\bar{y}^{\dot{\alpha}}\gg,F]+\frac{1}{3}\ll\partial_{\mu}\bar{\partial}_{\dot{\mu}}\mathrm{D}y\bar{y}F_{\alpha\alpha}y^{\alpha}y^{\alpha}\gg.\label{DF_eom}
\end{equation}

First, we process the first term on the r.h.s. (the second term will
be resolved by conjugation). Applying \eqref{d_C} to the problematic
factor yields
\begin{equation}
\ll\partial_{\mu}F_{\alpha\alpha}y^{\alpha}y^{\alpha}\gg=\partial_{\mu}F-\mathrm{D}_{\mu\dot{\alpha}}\bar{y}^{\dot{\alpha}}F+iy_{\mu}[\frac{1}{(N+1)(N+2)}\bar{F},F].\label{unf_d_Fyy}
\end{equation}
Therefore, one needs to process $\mathrm{D}_{\mu\dot{\alpha}}\bar{y}^{\dot{\alpha}}F$.
Contracting \eqref{DF_eom} with $\bar{y}^{\dot{\mu}}$ gives
\begin{equation}
\mathrm{D}_{\mu\dot{\mu}}\bar{y}^{\dot{\mu}}F=iy_{\mu}[\frac{1}{\bar{N}-1}\bar{F},F]+\frac{1}{3}\ll\partial_{\mu}\mathrm{D}y\bar{y}F_{\alpha\alpha}y^{\alpha}y^{\alpha}\gg.\label{D_ybar_F_initial}
\end{equation}
On the other hand, from \eqref{Dyy_F} and \eqref{d_C} one finds
\begin{equation}
\ll\partial_{\mu}\mathrm{D}y\bar{y}F_{\alpha\alpha}y^{\alpha}y^{\alpha}\gg=\partial_{\mu}\bar{N}F-\mathrm{D}_{\mu\dot{\alpha}}\bar{y}^{\dot{\alpha}}\bar{N}F+iy_{\mu}[\frac{1}{(N+1)(N+2)}\bar{F},\bar{N}F].\label{unf_d_Dyy_Fyy}
\end{equation}
Combining \eqref{D_ybar_F_initial} and \eqref{unf_d_Dyy_Fyy}, one
obtains after rearranging the Euler-operator ratios
\begin{equation}
\mathrm{D}_{\mu\dot{\alpha}}\bar{y}^{\dot{\alpha}}F=\frac{\bar{N}}{N+1}\partial_{\mu}F+iy_{\mu}\frac{2}{N+2}[\frac{1}{N+2}\bar{F},F]+iy_{\mu}[\frac{1}{(N+1)(N+2)}\bar{F},F].\label{D_ybar_F_final}
\end{equation}
Thus one finds for \eqref{unf_d_Fyy}
\begin{equation}
\ll\partial_{\mu}F_{\alpha\alpha}y^{\alpha}y^{\alpha}\gg=\frac{2}{N+1}(\partial_{\mu}F-iy_{\mu}[\frac{1}{N+2}\bar{F},F]).\label{1_term_final}
\end{equation}
Conjugation gives
\begin{equation}
\ll\bar{\partial}_{\dot{\mu}}\bar{F}_{\dot{\alpha}\dot{\alpha}}\bar{y}^{\dot{\alpha}}\bar{y}^{\dot{\alpha}}\gg=\frac{2}{\bar{N}+1}(\bar{\partial}_{\dot{\mu}}\bar{F}-i\bar{y}_{\dot{\mu}}[\frac{1}{\bar{N}+1}F,\bar{F}]).\label{2_term_final}
\end{equation}
These bring first two terms on the r.h.s. of \eqref{DF_eom} to the
admissible form.

Now we process the last term in \eqref{DF_eom}. By virtue of \eqref{d_C}
one has
\begin{eqnarray}
 &  & \ll\partial_{\mu}\bar{\partial}_{\dot{\mu}}\mathrm{D}y\bar{y}F_{\alpha\alpha}y^{\alpha}y^{\alpha}\gg=(\partial_{\mu}-\mathrm{D}_{\mu\dot{\alpha}}\bar{y}^{\dot{\alpha}})\ll\bar{\partial}_{\dot{\mu}}\mathrm{D}y\bar{y}F_{\alpha\alpha}y^{\alpha}y^{\alpha}\gg+\nonumber \\
 &  & +iy_{\mu}[\frac{1}{(N+1)(N+2)}\bar{F},\ll\bar{\partial}_{\dot{\mu}}\mathrm{D}y\bar{y}F_{\alpha\alpha}y^{\alpha}y^{\alpha}\gg].\label{d_d_Dyy_F}
\end{eqnarray}
Since obviously
\begin{equation}
\ll\bar{\partial}_{\dot{\mu}}F_{\alpha\alpha}y^{\alpha}y^{\alpha}\gg=0,
\end{equation}
one has from \eqref{dbar_C}
\begin{equation}
\mathrm{D}_{\alpha\dot{\mu}}y^{\alpha}F=\bar{\partial}_{\dot{\mu}}F+i\bar{y}_{\dot{\mu}}[\frac{1}{(\bar{N}+1)(\bar{N}+2)}F,F].
\end{equation}
Using this result and contracting \eqref{DF_eom} with $y^{\mu}$
yields
\begin{equation}
\ll\bar{\partial}_{\dot{\mu}}\mathrm{D}y\bar{y}F_{\alpha\alpha}y^{\alpha}y^{\alpha}\gg=\bar{\partial}_{\dot{\mu}}F-i\bar{y}_{\dot{\mu}}[\frac{1}{(\bar{N}+2)}F,F],\label{unf_dbar_Dyy_Fyy}
\end{equation}
which together with \eqref{D_ybar_F_final} turns \eqref{d_d_Dyy_F}
to
\begin{eqnarray}
 &  & \ll\partial_{\mu}\bar{\partial}_{\dot{\mu}}\mathrm{D}y\bar{y}F_{\alpha\alpha}y^{\alpha}y^{\alpha}\gg=\partial_{\mu}\bar{\partial}_{\dot{\mu}}F-i\bar{y}_{\dot{\mu}}\partial_{\mu}[\frac{1}{\bar{N}+2}F,F]+i[\frac{1}{(N+1)(N+2)}\bar{F},y_{\mu}\bar{\partial}_{\dot{\mu}}F]+\nonumber \\
 &  & +y_{\mu}\bar{y}_{\dot{\mu}}[\frac{1}{(N+1)(N+2)}\bar{F},[\frac{1}{\bar{N}+2}F,F]]+\mathrm{D}_{\mu\dot{\mu}}F-\frac{\bar{N}+1}{N+1}\partial_{\mu}\bar{\partial}_{\dot{\mu}}F-\nonumber \\
 &  & -2iy_{\mu}\bar{\partial}_{\dot{\mu}}\frac{1}{N+2}[\frac{1}{N+2}\bar{F},F]-iy_{\mu}\bar{\partial}_{\dot{\mu}}[\frac{1}{(N+1)(N+2)}\bar{F},F]+\nonumber \\
 &  & +i\bar{y}_{\dot{\mu}}[\frac{1}{\bar{N}+1}(\frac{\bar{N}}{N+1}\partial_{\mu}F+iy_{\mu}\frac{2}{N+2}[\frac{1}{N+2}\bar{F},F]+iy_{\mu}[\frac{1}{(N+1)(N+2)}\bar{F},F]),F]+\nonumber \\
 &  & +i\bar{y}_{\dot{\mu}}[\frac{1}{N}F,(\frac{\bar{N}}{N+1}\partial_{\mu}F+iy_{\mu}\frac{2}{N+2}[\frac{1}{N+2}\bar{F},F]+iy_{\mu}[\frac{1}{(N+1)(N+2)}\bar{F},F])].\label{3_term_final}
\end{eqnarray}
Substituting \eqref{1_term_final}, \eqref{2_term_final} and \eqref{3_term_final}
into \eqref{DF_eom} and combining like terms using the gauge algebra
Jacobi identity, one finally expresses the covariant derivative of
$F$ in $Y$-terms
\begin{eqnarray}
 &  & \mathrm{D}_{\mu\dot{\mu}}F=\frac{1}{N+1}\partial_{\mu}\bar{\partial}_{\dot{\mu}}F+iN[\frac{1}{N(N+1)}\bar{y}_{\dot{\mu}}\partial_{\mu}F,\frac{1}{N}F]-iy_{\mu}\bar{\partial}_{\dot{\mu}}\frac{1}{N+2}[\frac{1}{N+2}\bar{F},F]+\nonumber \\
 &  & +[\frac{i}{(N+1)(N+2)}y_{\mu}\bar{\partial}_{\dot{\mu}}\bar{F},F]+\frac{1}{2}y_{\mu}\bar{y}_{\dot{\mu}}[\frac{N+3}{(N+1)(N+2)}[\frac{1}{N+2}\bar{F},F],F]+\nonumber \\
 &  & +\frac{3}{2}y_{\mu}\bar{y}_{\dot{\mu}}[\frac{1}{(N+1)(N+2)}[\frac{1}{N}F,\bar{F}],F]+y_{\mu}\bar{y}_{\dot{\mu}}[\frac{1}{N+2}[\frac{1}{N+2}\bar{F},F],\frac{1}{N}F].\label{DF_final}
\end{eqnarray}

\subsubsection*{Poincaré symmetry and diffeomorphism-invariance}

The manifestly Poincaré-invariant relation \eqref{DF_final} is written
in Cartesian coordinates, together with \eqref{covar_derivative_def}
and \eqref{YM_field_postulate}, since they involve $\frac{\partial}{\partial x^{\mu\dot{\mu}}}$
inside of $\mathrm{D}_{\mu\dot{\mu}}$. The unfolded dynamics approach
requires manifest coordinate-independence, which is ensured by formulating
equations in terms of exterior forms. To attain this, we switch to
the fiber space picture: we claim that $F(Y|x)$ and $\bar{F}(Y|x)$
are now 0-forms on the Minkowski base manifold with some local coordinates
$x^{\underline{n}}$, while spinor variables $\{y^{\alpha},\bar{y}^{\dot{\alpha}}\}$
are coordinates in the fiber. This requires appropriate generalization
of $\mathrm{D}_{\mu\dot{\mu}}$ in \eqref{DF_final}.

According to the ideology of unfolding, global Poincaré symmetry of
a relativistic theory should arise in terms of the unique general
formula \eqref{unf_gauge_transf}. This is achieved by introducing
a non-dynamical 1-form $\Omega(x)$, which takes values in Lie algebra
$iso(1,3)$
\begin{equation}
\Omega=e^{\alpha\dot{\beta}}P_{\alpha\dot{\beta}}+\omega^{\alpha\alpha}M_{\alpha\alpha}+\bar{\omega}^{\dot{\alpha}\dot{\alpha}}\bar{M}_{\dot{\alpha}\dot{\alpha}},\label{Poincare_connection}
\end{equation}
with $P_{\alpha\dot{\alpha}}$, $M_{\alpha\alpha}$ and $\bar{M}_{\dot{\alpha}\dot{\alpha}}$
being generators of translations and rotations of $\mathbb{R}^{1,3}$,
and $e^{\alpha\dot{\beta}}$ and $\omega^{\alpha\alpha}$ ($\bar{\omega}^{\dot{\alpha}\dot{\alpha}}$)
being 1-forms of a vierbein and a Lorentz connection. 

$\Omega$ is subjected to the flatness condition (square brackets
stand for the $iso(1,3)$-commutator)
\begin{equation}
\mathrm{d}\Omega+\frac{1}{2}[\Omega,\Omega]=0,\label{flat_conn}
\end{equation}
so that the corresponding gauge symmetry \eqref{unf_gauge_transf}
is
\begin{equation}
\delta\Omega=\mathrm{d}\varepsilon(x)+[\Omega,\varepsilon]
\end{equation}
and describes an infinite-dimensional freedom in switching between
all possible local coordinates on $\mathbb{R}^{1,3}$. This boils
down to 10-dimensional global Poincaré symmetry after fixing some
particular solution $\Omega_{0}$ and restricting to those residual
$\varepsilon(x)$ which leave it invariant
\begin{equation}
\mathrm{d}\varepsilon_{0}+[\Omega_{0},\varepsilon_{0}]=0.\label{glob_symm}
\end{equation}

The simplest non-degenerate global solution to \eqref{flat_conn}
is provided by Cartesian coordinates
\begin{equation}
e_{\underline{m}}{}^{\alpha\dot{\beta}}=(\bar{\sigma}_{\underline{m}})^{\dot{\beta}\alpha},\quad\omega_{\underline{m}}{}^{\alpha\alpha}=0,\quad\bar{\omega}_{\underline{m}}{}^{\dot{\alpha}\dot{\alpha}}=0,\label{cartes_coord}
\end{equation}
with global symmetries \eqref{glob_symm} parameterized by $x$-independent
$\xi^{\alpha\dot{\beta}}$, $\xi^{\alpha\alpha}$ and $\bar{\xi}^{\dot{\alpha}\dot{\alpha}}$
\begin{equation}
\varepsilon_{0}^{\alpha\dot{\beta}}=\xi^{\alpha\dot{\beta}}+\xi^{\alpha}\text{}_{\gamma}(\bar{\sigma}_{\underline{m}})^{\dot{\beta}\gamma}x^{\underline{m}}+\bar{\xi}^{\dot{\beta}}\text{}_{\dot{\gamma}}(\bar{\sigma}_{\underline{m}})^{\dot{\gamma}\alpha}x^{\underline{m}},\quad\varepsilon_{0}^{\alpha\alpha}=\xi^{\alpha\alpha},\quad\bar{\varepsilon}_{0}^{\dot{\alpha}\dot{\alpha}}=\bar{\xi}^{\dot{\alpha}\dot{\alpha}}.\label{glob_cart_sol}
\end{equation}

Analogously, in order to realize the Yang\textendash Mills gauge symmetry
via \eqref{unf_gauge_transf}, one introduces a 1-form $A(x)$, with
$A_{\alpha\dot{\alpha}}(x)$ being its expansion in the vierbein 
\begin{equation}
A(x)=e^{\alpha\dot{\alpha}}A_{\alpha\dot{\alpha}}.
\end{equation}
Now an appropriate coordinate-independent generalization of $\mathrm{D}_{\mu\dot{\mu}}$
is a 1-form operator $\mathrm{D}$, supplemented by Lorentz-connection
terms rotating fiber coordinates $Y$,
\begin{equation}
\mathrm{D}:=\mathrm{d}+\omega^{\alpha\alpha}y_{\alpha}\partial_{\alpha}+\bar{\omega}^{\dot{\alpha}\dot{\alpha}}\bar{y}_{\dot{\alpha}}\bar{\partial}_{\dot{\alpha}}-i[A,\bullet].
\end{equation}
In Cartesian coordinates \eqref{cartes_coord}, this indeed boils
down to
\begin{equation}
\mathrm{D}=\mathrm{d}x^{\mu\dot{\mu}}\mathrm{D}_{\mu\dot{\mu}}.
\end{equation}

\subsubsection*{Unfolded Yang\textendash Mills equations and unfolding maps}

Now we are ready to write down an unfolded system for Yang\textendash Mills
theory, which is the main result of the note. Contracting \eqref{YM_tensor}
and \eqref{DF_final} with the vierbeins yields
\begin{eqnarray}
 &  & \mathrm{d}A+[A,A]=\frac{1}{4}e^{\alpha}{}_{\dot{\beta}}e^{\alpha\dot{\beta}}\partial_{\alpha}\partial_{\alpha}F|_{\bar{y}=0}+\frac{1}{4}e_{\beta}{}^{\dot{\alpha}}e^{\beta\dot{\alpha}}\bar{\partial}_{\dot{\alpha}}\bar{\partial}_{\dot{\alpha}}\bar{F}|_{y=0},\label{A_unf_eq}\\
 &  & \mathrm{D}F=\frac{1}{N+1}e\partial\bar{\partial}F+iN[\frac{1}{N(N+1)}e\partial\bar{y}F,\frac{1}{N}F]-iey\bar{\partial}\frac{1}{N+2}[\frac{1}{N+2}\bar{F},F]+\nonumber \\
 &  & +[\frac{i}{(N+1)(N+2)}ey\bar{\partial}\bar{F},F]+\frac{1}{2}ey\bar{y}[\frac{N+3}{(N+1)(N+2)}[\frac{1}{N+2}\bar{F},F],F]+\nonumber \\
 &  & +\frac{3}{2}ey\bar{y}[\frac{1}{(N+1)(N+2)}[\frac{1}{N}F,\bar{F}],F]+ey\bar{y}[\frac{1}{N+2}[\frac{1}{N+2}\bar{F},F],\frac{1}{N}F],\label{F_unf_eq}
\end{eqnarray}
plus a conjugate equation for $\bar{F}$ resulting from exchanging
barred and unbarred objects in \eqref{F_unf_eq}. Here
\begin{equation}
e\partial\bar{\partial}:=e^{\alpha\dot{\beta}}\partial_{\alpha}\bar{\partial}_{\dot{\beta}},\quad e\partial\bar{y}:=e^{\alpha\dot{\beta}}\partial_{\alpha}\bar{y}_{\dot{\beta}},\quad ey\bar{\partial}:=e^{\alpha\dot{\beta}}y_{\alpha}\bar{\partial}_{\dot{\beta}}\quad ey\bar{y}:=e^{\alpha\dot{\beta}}y_{\alpha}\bar{y}_{\dot{\beta}}.
\end{equation}

A full spectrum of unfolded fields consists of a 1-form $\Omega$
describing Minkowski background, a 1-form of the gauge potential $A$
and the 0-forms of master-fields $F(Y|x)$ and $\bar{F}(Y|x)$ encoding
the Yang\textendash Mills tensor together with an infinite tower of
its covariant derivatives. The corresponding unfolded equations are
\eqref{flat_conn}, \eqref{A_unf_eq} and \eqref{F_unf_eq} plus a
conjugate for $\bar{F}$. The formulation is manifestly diffeomorphism-invariant.
1-forms $\Omega$ and $A$ give rise to two manifest symmetries in
accordance with \eqref{unf_gauge_transf}: the global (after fixing
$\Omega$) Poincaré one and the local Yang\textendash Mills one. The
Yang\textendash Mills symmetry is realized as
\begin{equation}
\delta A(x)=\mathrm{D}\varepsilon(x),\quad\delta F(Y|x)=i[\varepsilon(x),F(Y|x)],\quad\delta\bar{F}(Y|x)=i[\varepsilon(x),\bar{F}(Y|x)].\label{YM_gauge_transform}
\end{equation}

By construction, the unfolded system is consistent, provided unfolded
0-form fields obey the helicity constraint \eqref{helicity}. At the
same time, a direct check of \eqref{unf_consist} seems hardly executable
due to the complexity of the equations. Note that the system is at
most cubic in master-field 0-forms, which is surprising on its own.
One of the possible forms of solution to the system is \eqref{YM_field_postulate}
with primary fields subjected to \eqref{YM_eq_F}. Let us quickly
derive this directly from \eqref{F_unf_eq}.

Master-fields $F$ and $\bar{F}$ are assumed to be analytical in
$Y$ and obey \eqref{helicity}. For the sake of brevity, here and
below we consider only an anti-selfdual component $F$, but everything
applies to $\bar{F}$ as well. Acting on \eqref{F_unf_eq} with $y^{\beta}\bar{y}^{\dot{\beta}}\frac{\delta}{\delta e^{\beta\dot{\beta}}}$
produces \eqref{Dyy_F}, whose solution, accounting for \eqref{helicity},
is \eqref{YM_field_postulate}. On the other hand, acting on \eqref{F_unf_eq}
with $\frac{1}{2}\partial_{\mu}\partial^{\beta}\frac{\delta}{\delta e^{\beta\dot{\mu}}}$
and putting $\bar{y}^{\dot{\alpha}}=0$ yields, accounting for \eqref{helicity}
again,
\begin{equation}
\mathrm{D}_{\beta\dot{\mu}}F^{\beta}{}_{\mu}=0.
\end{equation}
Thus, the unfolded system presented above indeed provides a consistent
manifestly diffeomorphism- and gauge-invariant first-order formulation
of $4d$ Yang-Mills theory. Unfolded master-fields $F(Y|x)$ and $\bar{F}(Y|x)$
encode all on-shell d.o.f. of the Yang-Mills tensor as expansions
in auxiliary spinors $Y$.

The system allows for two obvious reductions. The first one is the
anti-selfdual case $\bar{F}(Y|x)=0$ (or the selfdual $F(Y|X)=0$),
then only first two terms on the r.h.s. of \eqref{F_unf_eq} survive.
In different (tensor) terms, this was presented in \cite{sdym}. The
second one is the abelian case with all commutators vanishing, then
only the first term on the r.h.s. of \eqref{F_unf_eq} remains.

One can think of the equation \eqref{F_unf_eq} as defining an unfolding
map from $x$-space to $Y$-space
\begin{equation}
F_{\alpha\alpha}(x)|_{on-shell}\rightarrow F(Y|x)\rightarrow\mathcal{F}(Y):=F(Y|x=0).\label{unf_map}
\end{equation}
Then \eqref{YM_field_postulate} explicitly realizes the first arrow
in \eqref{unf_map}. The field $\mathcal{F}(Y)$ carries precisely
the same information as on-shell $F_{\alpha\alpha}$ does. In a sense,
spinors $Y=\{y^{\alpha};\bar{y}^{\dot{\alpha}}\}$ effectively replace
space-time coordinates $x^{\alpha\dot{\alpha}}$ for on-shell configurations,
hence being conjugate to spinor-helicity variables that resolve light-like
momenta $p^{\alpha\dot{\alpha}}=\pi^{\alpha}\bar{\pi}^{\dot{\alpha}}$.
This is the way the unfolded system imposes e.o.m. on primary fields:
via \eqref{unf_map}, it maps $4d$ space-time fields onto an effectively
$3d$ hypersurface (in the sense that $y^{\alpha}\bar{y}^{\dot{\alpha}}$
is a light-like vector). Note however, that the role of auxiliary
spinors is much more important and sophisticated. As follows directly
from \eqref{YM_field_postulate}, they in fact equalize and mix translational
and spin degrees of freedom. So they should not be thought of simply
as coordinates on some null hypersurface in $4d$ Minkowski space. 

To get a better idea of the unfolding map \eqref{unf_map}, it is
instructive to consider the abelian case, where the unfolding exponent
in \eqref{YM_field_postulate} boils down to a space-time translation
so that \eqref{unf_map} can be constructed explicitly. Then a plane-wave
solution to \eqref{A_unf_eq}, \eqref{F_unf_eq} in Cartesian coordinates
is
\begin{equation}
A_{\alpha\dot{\alpha}}(x)=\pi_{\alpha}\bar{\mu}_{\dot{\alpha}}e^{i\pi_{\beta}\bar{\pi}_{\dot{\beta}}x^{\beta\dot{\beta}}}+c.c.,\hspace*{1em}F(Y|x)=i\bar{\pi}_{\dot{\alpha}}\bar{\mu}^{\dot{\alpha}}(\pi_{\alpha}y^{\alpha})^{2}e^{i\pi_{\beta}\bar{\pi}_{\dot{\beta}}(x^{\beta\dot{\beta}}+y^{\beta}\bar{y}^{\dot{\beta}})}
\end{equation}
with $\bar{\mu}_{\dot{\alpha}}$ being an arbitrary reference spinor,
defined up to a gauge transformation $\bar{\mu}_{\dot{\alpha}}\rightarrow\bar{\mu}_{\dot{\alpha}}+const\cdot\bar{\pi}_{\dot{\alpha}}$.
Putting $x=0$, one has
\begin{equation}
\mathcal{F}(Y)=i\bar{\pi}_{\dot{\alpha}}\bar{\mu}^{\dot{\alpha}}(\pi_{\alpha}y^{\alpha})^{2}e^{i\pi_{\beta}y^{\beta}\bar{\pi}_{\dot{\beta}}\bar{y}^{\dot{\beta}}},\label{F_Y_plane_wave}
\end{equation}
which represents a plane-wave Maxwell tensor formulated purely in
$Y$-terms.

In fact, one can start directly from \eqref{F_Y_plane_wave} and then
make use of \eqref{F_unf_eq} in order to fully recover $x$-dependence.
This implies that although we have derived the unfolded equation \eqref{F_unf_eq}
starting from postulating \eqref{YM_field_postulate}, for \eqref{F_unf_eq}
the expression \eqref{YM_field_postulate} \emph{per se} is nothing
more than just one of many possible forms of solution. In particular,
instead of \eqref{unf_map} one can think of \eqref{F_unf_eq} as
defining a $Y$-to-$x$ map (a similar interpretation has been proposed
in \cite{unf_penrose})
\begin{equation}
\mathcal{F}(Y)\rightarrow F(Y|x)\rightarrow F_{\alpha\alpha}(x)|_{on-shell}.\label{unf_map_reverse}
\end{equation}
In this picture, $\mathcal{F}(Y)$ is completely unconstrained aside
from the helicity condition
\begin{equation}
(N-\bar{N})\mathcal{F}(Y)=2\mathcal{F}(Y)\hspace{1em}\Rightarrow\hspace{1em}\mathcal{F}(Y)=\mathcal{F}_{\alpha\alpha}(y\bar{y})y^{\alpha}y^{\alpha}.
\end{equation}
This distinguishes the unfolded formulation constructed here from
that of \cite{ActionsCharges}, where the fiber structure is tantamount
to the base one. In the abelian case, the map \eqref{unf_map_reverse}
is obviously realized as
\begin{equation}
F(Y|x)=\exp(\frac{1}{N+1}x^{\beta\dot{\beta}}\partial_{\beta}\bar{\partial}_{\dot{\beta}})\mathcal{F}_{\alpha\alpha}(y\bar{y})y^{\alpha}y^{\alpha}.\label{x_recover}
\end{equation}
It generates a solution to \eqref{F_unf_eq}, and hence implicitly
to Maxwell equations, for arbitrary $\mathcal{F}_{\alpha\alpha}(y\bar{y})$.
This indicates that relativistic dynamics in fact can be realized
in terms of $Y$ without any reference to a space-time. The corresponding
action principle for an arbitrary-mass integer-spin field was constructed
in \cite{spinor_action} by means of supplementing the set of $Y$
with a Lorentz-invariant proper-time coordinate serving as an evolution
parameter.

From this perspective, Yang\textendash Mills theory is characterized
by the unconstrained field $\mathcal{F}(Y)=\mathcal{F}_{\alpha\alpha}(y\bar{y})y^{\alpha}y^{\alpha}$
and its conjugate $\bar{\mathcal{F}}$. These field form representations
of both the gauge and Poincaré algebras, as defined by the unfolded
system via the general formula \eqref{unf_gauge_transf}. At first
glance, this might suggest the theory is already solved, since no
equations constrain $\mathcal{F}$ and $\bar{\mathcal{F}}$. However,
the key complication preserving the theory's non-triviality lies in
the nonlinear realization of Poincaré translations in $\mathcal{F}$
and $\bar{\mathcal{F}}$, as evident from \eqref{F_unf_eq}. Considering
this, the unfolding map \eqref{unf_map_reverse} suggests that the
unfolded dynamics approach may provide a new way to study the problem
of integrability of classical e.o.m. and constructing exact solutions.
In particular, the question is whether it is possible to extend \eqref{x_recover}
to the non-abelian case. If so, an appropriate unfolding map will
generate solutions to Yang\textendash Mills equations in Minkowski
space. Additionally, the unfolded framework provides a systematic
method for identifying theory invariants: all gauge-invariant conserved
charges are classified through the cohomology of a certain operator
determined by the unfolded equations \cite{ActionsCharges}.

Another interesting problem is to relate the unfolded dynamics approach
to twistor theory \cite{twistor1,twistor2}. Potentially, twistors
may arise from treating an unfolded system as defining an unfolding
map to $x$-space from some complex plane in $(Y|x)$-space, different
from $x=0$ of \eqref{unf_map_reverse} and associated with the incidence
relation. Then this complex plane should be identified with the twistor
space, and the corresponding unfolding map with the Penrose transform.
A related discussion of these questions can be found in \cite{twistor_unf}.

\subsubsection*{Conclusion\label{SEC_conclusions}}

In this note, we constructed an unfolded formulation of $4d$ pure
Yang\textendash Mills theory making use of the unfolding method proposed
in \cite{qed} and improved here by the preliminary derivation of
general relations for unfolded functions.

A natural generalization of our result would be the inclusion of charged
matter. It is straightforward and should not present any difficulties,
as shown by the example of scalar electrodynamics \cite{qed}.

Another possible direction of further research is to include supersymmetry,
as well as to manifest conformal symmetry. To this end one needs to
introduce the corresponding gauge 1-forms of (super)conformal gravity
in addition to the Poincaré connection and to deform appropriately
the unfolded equations. In its turn, this requires further non-trivial
modifications of the unfolding method of \cite{qed}. In particular,
it would be interesting to unfold $\mathcal{N}=4$ super-Yang\textendash Mills
theory in order to apply unfolding tools to the problems of $AdS/CFT$
and amplitudes. Another interesting problem is to unfold Einstein
equations.

Quantization of the unfolded Yang\textendash Mills system can be performed
along the lines of \cite{unfQFT} but with necessary modifications
in order to include ghosts. The first step is to build an off-shell
extension of the unfolded system constructed here, which is equivalent
to coupling it to external currents \cite{off-shell1,off-shell2}.
After that, this off-shell system can be elevated to an unfolded Schwinger\textendash Dyson
system for correlation functions of quantum unfolded fields.

Finally, there are two vague but potentially promising topics: the
study of integrability of classical e.o.m. and the derivation of twistors.
Both of them seem to be related to the investigation of various unfolding
maps defined by the unfolded Yang\textendash Mills equations.

\section*{Acknowledgments}

The author is grateful to A.V. Korybut, D.S. Ponomarev and M.A. Vasiliev
for valuable comments and remarks.

\end{document}